\documentclass[superscriptaddress,twocolumn,aps,pra,10pt]{revtex4-2}

\usepackage{graphicx}
\usepackage{amsmath,amssymb,mathtools}

\usepackage{kotex}
\usepackage[T1]{fontenc}

\usepackage[caption=false]{subfig}

\usepackage{xcolor}

\usepackage[colorlinks=true,linkcolor=blue,citecolor=blue,urlcolor=blue]{hyperref}

\begin{document}


\title{Jacobi--Anger Density Estimation for Energy Distribution of Quantum States}

\author{Kyeongan Park}
\affiliation{Department of Chemistry, Sungkyunkwan University, Suwon 16419, Republic of Korea}

\author{Gwonhak Lee}
\affiliation{SKKU Advanced Institute of Nanotechnology (SAINT),\\
Sungkyunkwan University , Suwon 16419, Republic of Korea}

\author{Minhyeok Kang}
\affiliation{SKKU Advanced Institute of Nanotechnology (SAINT),\\
Sungkyunkwan University , Suwon 16419, Republic of Korea}

\author{Youngjun Park}
\affiliation{Department of Chemistry, Yonsei University, Seoul 03722, Republic of Korea}

\author{Joonsuk Huh}
\email{joonsukhuh@yonsei.ac.kr}
\affiliation{Department of Chemistry, Yonsei University, Seoul 03722, Republic of Korea}
\affiliation{Department of Quantum Information, Yonsei University, Incheon 21983, Republic of Korea}

\begin{abstract}
The energy distribution of a quantum state is essential for accurately estimating a molecule’s ground state energy in quantum computing. Directly obtaining this distribution requires full Hamiltonian diagonalization, which is computationally prohibitive for large-scale systems. A more practical strategy is to approximate the distribution from a finite set of Hamiltonian moments. However, reconstructing an accurate distribution from only a limited number of moments remains a significant challenge. In this work, we introduce Jacobi--Anger Density Estimation (JADE), a non-parametric, quantum-inspired method designed to overcome this difficulty. JADE reconstructs the characteristic function from a finite set of moments using the Jacobi--Anger expansion and then estimates the underlying distribution via an inverse Fourier transform. We demonstrate that JADE can accurately recover the energy distribution of a quantum state for a molecular system. Beyond quantum chemistry, we also show that JADE is broadly applicable to the estimation of complicated probability density functions in various other scientific and engineering fields. Our results highlight JADE as a powerful and versatile tool for practical quantum systems, with the potential to significantly enhance ground state energy estimation and related applications.
\end{abstract}

\maketitle

\section{Introduction}
\label{sec:intro}

In quantum computing, estimating the ground-state energies of molecules is an important goal \cite{9_Initial}. Many of the most promising algorithms for this task, such as Quantum Phase Estimation (QPE), the Variational Quantum Eigensolver (VQE), and Krylov Quantum Diagonalization (KQD), have performance that strongly depends on the quality of the chosen initial state~\cite{45_initial_prep_important}. A high-quality initial state is crucial because it significantly influences the algorithm's efficiency and success: for instance, it ensures faster convergence in variational methods like VQE~\cite{51_VQE&QPE, 55_VQE}, increases the success probability in estimation algorithms in QPE, which can reduce the required runtime or circuit repetitions~\cite{51_VQE&QPE, 52_QPE, 56_QPE, 57_QPE}, and admits convergence guarantees for KQD under a sufficient ground-state overlap, which is often required in practice to attain efficient performance~\cite{53_KQD, 58_KQD}.

A good initial state is one with high overlap with the true ground state. However, evaluating this overlap directly is impractical, as it requires prior knowledge of the true ground state one seeks to find. To evaluate candidate initial states in practice, the concept of a state’s energy distribution has been introduced~\cite{9_Initial}. This offers a more practical approach, since the distribution is a property of the candidate state itself, rather than a comparison to the unknown ground state. This distribution describes the probability of measuring different energy eigenvalues from the state. Accordingly, the distribution can serve as a pre-screening step: by assessing whether it has appreciable probability mass below the best energy obtained by classical methods, one can determine whether QPE or other methods can be expected to improve upon that value. From this perspective, a good initial state is one with a high probability density in the low-energy region of its distribution. By computing the energy distribution for each candidate, we can compare them and select the most promising. Accurate estimation of the energy distribution is therefore essential.

Unfortunately, obtaining the exact energy distribution requires full diagonalization of the Hamiltonian, which is computationally prohibitive. By contrast, computing the Hamiltonian moments is relatively inexpensive. Thus, being able to reconstruct the energy distribution using only the moments offers a highly efficient solution.

Several conventional methods attempt to estimate a probability density function (PDF) from a finite set of moments (or cumulants), including the Gram--Charlier and Edgeworth expansions~\cite{11_gram_2,blinnikov1998expansions}. These approaches approximate the PDF by adding correction terms to a Gaussian reference distribution. While they avoid expensive computations such as nonlinear optimization, they cannot capture distributions that deviate strongly from Gaussianity, such as multimodal energy distributions. Consequently, they are not well suited for quantum states~\cite{4_gram_1, 9_Initial}.

As an alternative, quantum algorithms have been proposed for estimating energy distribution, such as coarse-QPE~\cite{9_Initial}. While promising, such approaches still face significant practical limitations. Coarse-QPE reduces the number of controlled-unitary operations required by full QPE, thereby lowering the overall resource cost. Nevertheless, these controlled operations remain an unavoidable bottleneck: they are resource-intensive and error-prone on today’s quantum hardware. This makes the practical deployment of coarse-QPE highly challenging.

Here, we present Jacobi--Anger Density Estimation (JADE), a new moment-based method for density estimation. JADE requires only the Hamiltonian moments, avoids complex optimization routines, and accurately estimates even multimodal, non-Gaussian energy distributions. It does not rely on corrections to a Gaussian distribution like expansion-based methods and, in contrast to coarse-QPE, is efficiently executable on classical hardware.

The method is based on a structural analogy between two functions: the quantum autocorrelation function $\langle \psi | \mathrm{e}^{-\mathrm{i} \hat{H} t} | \psi \rangle$, which encodes the energy distribution of a quantum state $| \psi \rangle$ under the evolution of a Hamiltonian $\hat{H}$, and the characteristic function $\langle \mathrm{e}^{\mathrm{i}tX} \rangle_X$, which contains the full information of a distribution for a random variable $X$. Exploiting this analogy, JADE reconstructs the characteristic function from a finite set of moments using the Jacobi--Anger expansion, and then recovers the energy distribution via an inverse Fourier transform.

Beyond quantum chemistry, JADE can be applied more broadly as a general PDF estimation method. JADE differentiates itself from the maximum entropy method (MEM)~\cite{7_MEM_1, 8_MEM_2, 14_MEM_unbias, 54_MEM_unbias_2} by not requiring complex nonlinear optimization. Additionally, it does not rely on sample data, kernel selection, or bandwidth tuning like kernel density estimation (KDE)~\cite{5_KDE_1,6_KDE_2,16_KDE_3}. While existing methods suffer from either limited accuracy or excessive computational overhead, JADE provides highly accurate estimates, even for complicated distributions, through a simple closed-form expression. In this sense, JADE is a non-parametric, quantum-inspired method for efficiently estimating PDFs from finite moments.

\begin{figure}[t!]
  \centering
  \includegraphics[width=1.0\linewidth]{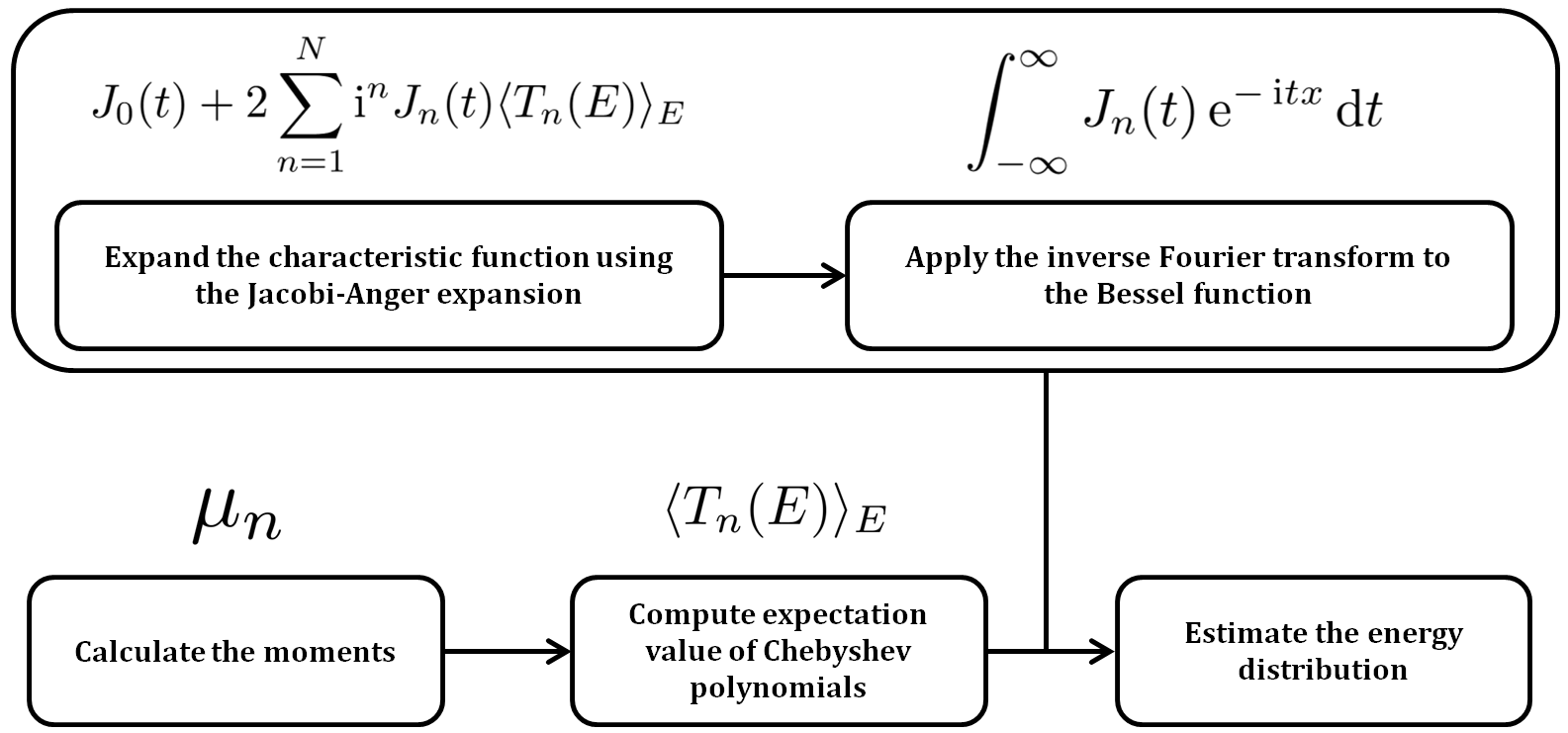}
  \caption{
  Schematic overview of the proposed JADE method. The process begins with computing the distribution’s moments and the corresponding expectation values of Chebyshev polynomials. These are then incorporated into the Jacobi--Anger expansion to approximate the characteristic function. Finally, the energy distribution is obtained analytically through the inverse Fourier transform.}
  \label{fig:1_Scheme}
\end{figure}

The overall workflow of JADE is illustrated in Fig.~\ref{fig:1_Scheme}. Starting from a finite set of moments, the method constructs an approximation of the characteristic function via the Jacobi--Anger expansion. Then it applies an inverse Fourier transform to estimate the energy distribution. This simple, non-iterative procedure provides a significant advantage, as discussed in the following sections.

The remainder of this paper is organized as follows. Section~\ref{sec:Methods} introduces JADE in detail. Section~\ref{sec:Demonstration} demonstrates its performance by estimating molecular energy distributions and benchmarking against alternative methods on complex PDFs. Section~\ref{sec:Conclusion} summarizes our findings and outlines potential applications in quantum computing. A detailed derivation of JADE is provided in Appendix~\ref{sec:Ap_A}, Appendix~\ref{sec:Ap_C} proves that JADE coincides with the optimal solution under a specifically defined weighted $L_2$ loss, which is detailed in Section~\ref{sec:Methods}.

\section{Methods}
\label{sec:Methods}

In the introduction, we highlighted the importance of energy distributions and introduced JADE as a method for efficiently estimating the energy distributions of quantum states. We also noted that JADE is not limited to energy distributions but can also be applied to a wide variety of PDFs. In this section, we present the details of JADE, focusing specifically on how it estimates the energy distribution of a quantum state from a finite number of moments.
JADE builds on the Fourier transform relationship between a PDF and its characteristic function~\cite{11_gram_2,12_fourier}. In the context of quantum states, the relevant PDF is the energy distribution $P(\epsilon)$. Using this connection, JADE reconstructs the characteristic function from a finite set of moments via the Jacobi--Anger expansion. This section begins with a brief review of moments and characteristic functions, which form the foundation of JADE.
The $n$th moment $\mu_n$ of the energy distribution $P(\epsilon)$ is defined as
\begin{equation}
\label{energy_moment}
\mu_n \coloneqq \int_{-\infty}^{\infty}{\epsilon^n P(\epsilon)\, \mathrm{d}\epsilon},
\end{equation}
where $\epsilon$ is an energy value and $n$ is a non-negative integer.
 The energy distribution of a quantum state $\vert \Psi\rangle$ is resolved with eigenenergies ${\epsilon_k}$ of eigenstates ${\vert\phi_{k}\rangle}$ as 
\begin{equation}
P(\epsilon)=\sum_{k}\vert \gamma_{k}\vert^{2}K_\sigma(\epsilon-\epsilon_{k})
,
\end{equation}
where $\vert \Psi\rangle=\sum_{k}\gamma_{k}\vert\phi_{k}\rangle$ , $\gamma_k = \langle \phi_k \vert \Psi \rangle$ and $K_\sigma$ is the kernel function (such as a Gaussian or Lorentzian), where the parameter $\sigma$ determines the broadening width. In the limit $\sigma \to 0$, $K_\sigma$ approaches the Dirac delta function, recovering the exact discrete probabilities $|\gamma_k|^2$ at each eigenstate ${\vert\phi_{k}\rangle}$. Equivalently, the $n$th moment $\mu_n$ can be written as the expectation value of the $n$th power of the Hamiltonian $\hat{H}$:
\begin{equation}
\label{energy_hamiltonian_moment}
\mu_n \coloneqq \langle \Psi | {\hat{H}}^n | \Psi \rangle
= \langle E^n \rangle_E.
\end{equation}

The characteristic function of a random variable for energy $E$ is then defined as~\cite{36_characteristic}:
\begin{equation}
\label{energy_characteristic}
\varphi_E(t) = \langle \mathrm{e}^{\mathrm{i}tE} \rangle_E = \int_{-\infty}^{\infty}{\mathrm{e}^{\mathrm{i}t\epsilon} P(\epsilon)\, \mathrm{d}\epsilon},
\end{equation}
where $t$ is a real variable. Using the Jacobi--Anger expansion~\cite{20_Jacobi-anger_expansion}, this characteristic function can be expressed as
\begin{align}
    \varphi_E(t)
    &= J_0(t) + 2 \sum_{n=1}^{\infty}{\mathrm{i}^n J_n(t)\langle T_n(E) \rangle}_E \label{energy_Jacobi-Anger_exact} \\
    &\approx J_0(t) + 2 \sum_{n=1}^{N}{\mathrm{i}^n J_n(t)\langle T_n(E) \rangle}_E
    \label{energy_Jacobi-Anger_approx}
    \eqqcolon \widetilde{\varphi}_E(t)
    ,
\end{align}
where $\widetilde{\varphi}_{E}(t)$ is the approximate characteristic function, obtained by truncating the expansion at degree $N$.

Here $J_n(t)$ is the Bessel function of the first kind of order $n$, defined as~\cite{32_bessel}:
\begin{equation}
\label{eq:5_Bessel_def}
J_n(t) = \sum_{m=0}^{\infty} \frac{(-1)^m}{m! \ \Gamma(m+n+1)}
\Big( \frac{t}{2} \Big)^{2m+n},
\end{equation}
where $\Gamma(z)$ denotes the gamma function, and $T_n(E)$ denotes the Chebyshev polynomial of the first kind of degree $n$ for energy $E$, defined by~\cite{19_chev}:
\begin{equation}
\label{eq:6_Chebyshev_def}
T_n(E) = \cos{(n \, \arccos{(E)})}.
\end{equation}
The Chebyshev polynomials satisfy $\left\vert T_n(E) \right\vert \leq 1$ for $E \in [-1, 1]$.

By combining Eqs.~\eqref{energy_moment}, \eqref{energy_hamiltonian_moment}, and \eqref{eq:6_Chebyshev_def}, the expectation value $\langle T_n(E) \rangle_E$ can be written as a linear combination of moments: $\langle T_n(E) \rangle_E=\sum_{m=0}^n c_{n,m} \langle E^m \rangle_E$, where $c_{n,m}$ denotes the Chebyshev coefficient of $m$th monomial.
Finally, by applying the inverse Fourier transform to $\widetilde{\varphi}_{E}(t)$ (Eq.~\eqref{energy_characteristic}), we obtain
\begin{align}
    \nonumber
    P(\epsilon)
    &\approx
    \frac{1}{2\pi} \int_{-\infty}^{\infty}{\mathrm{e}^{-\mathrm{i}t\epsilon} \ \widetilde{\varphi}_E(t)\, \mathrm{d}t} \nonumber
    \\
    &=
    \frac{1}{\pi}
    \Bigg[
    \frac{\langle T_0(E) \rangle_E T_0(\epsilon)}{\sqrt{1-\epsilon^2}}
    +
    \sum_{n=1}^N
    \frac{2\langle T_n(E) \rangle_E T_n(\epsilon)}{\sqrt{1-\epsilon^2}}
    \Bigg].
    \label{energy_JADE_closed}
\end{align}

Eq.~\eqref{energy_JADE_closed} provides a closed-form expression for the energy distribution via JADE. Despite its non-parametric property and simplicity—requiring only the distribution’s moments as input—this expression yields highly accurate results, as will be demonstrated in Section~\ref{sec:Demonstration}. A generalized derivation beyond energy distributions is presented in Appendix~\ref{sec:Ap_A}.

The high accuracy of this expression is not incidental: it is a consequence of the method's mathematical optimality. Specifically, Eq.~\eqref{energy_JADE_closed} is equivalent to the solution that minimizes a weighted $L_2$ loss. This loss is defined for the function space spanned by the JADE's basis, where the inner product between two functions, $F(\epsilon)$ and $G(\epsilon)$, is given by
\begin{equation}
\label{eq:Inner_product_space}
\langle F, G \rangle_w := \displaystyle \int_{-1} ^{1} F(\epsilon)\overline{G(\epsilon)}\sqrt{1-\epsilon^2}\mathrm{d}\epsilon.
\end{equation}
A detailed proof of this optimality is provided in Appendix~\ref{sec:Ap_C}.

To summarize, JADE proceeds as follows (see also Fig.~\ref{fig:1_Scheme}): compute the distribution’s moments $\mu_n$; evaluate the expectation values of the Chebyshev polynomials $\langle T_n(E)\rangle_E$; substitute these into the Jacobi--Anger expansion to obtain the approximate characteristic function $\widetilde{\varphi}_E(t)$; and finally, apply the analytic inverse Fourier transform to estimate the energy distribution.

\section{Demonstration}
\label{sec:Demonstration}
In Section~\ref{sec:Methods}, we outlined the derivation of JADE and presented its compact closed-form expression in Eq.~\eqref{energy_JADE_closed}. In this section, we first apply JADE to estimate the energy distributions of molecular quantum states and discuss its potential applications in quantum computing. We then extend the analysis beyond quantum systems, demonstrating—as highlighted in the introduction—that JADE also performs effectively on complicated distributions that conventional methods such as the Gram--Charlier expansion, MEM, and KDE fail to capture. To illustrate this versatility, we apply JADE to four representative types of distributions.

\subsection{Energy distribution}
\label{sec:energy_distribution}

\begin{figure}[t!]
  \centering
  \begin{minipage}[b]{\linewidth}
    \subfloat[Gram--Charlier expansion]{%
      \includegraphics[width=\linewidth]{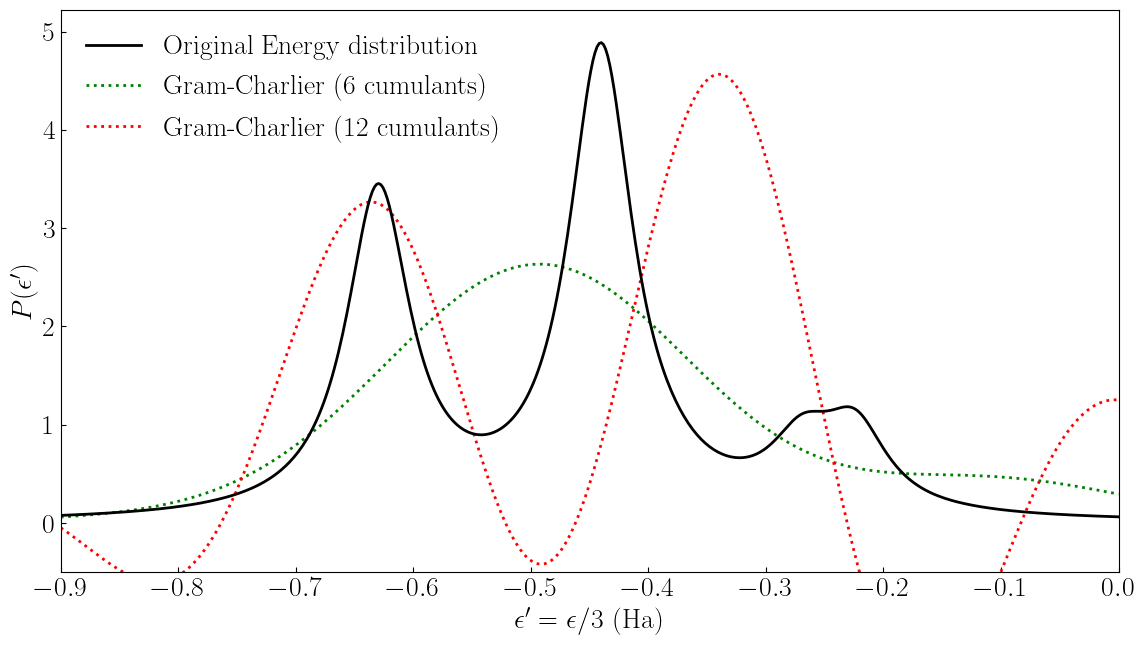}%
      \label{fig:energy_distribution_GC}%
    }
  \end{minipage}
  \begin{minipage}[b]{\linewidth}
    \subfloat[JADE]{%
      \includegraphics[width=\linewidth]{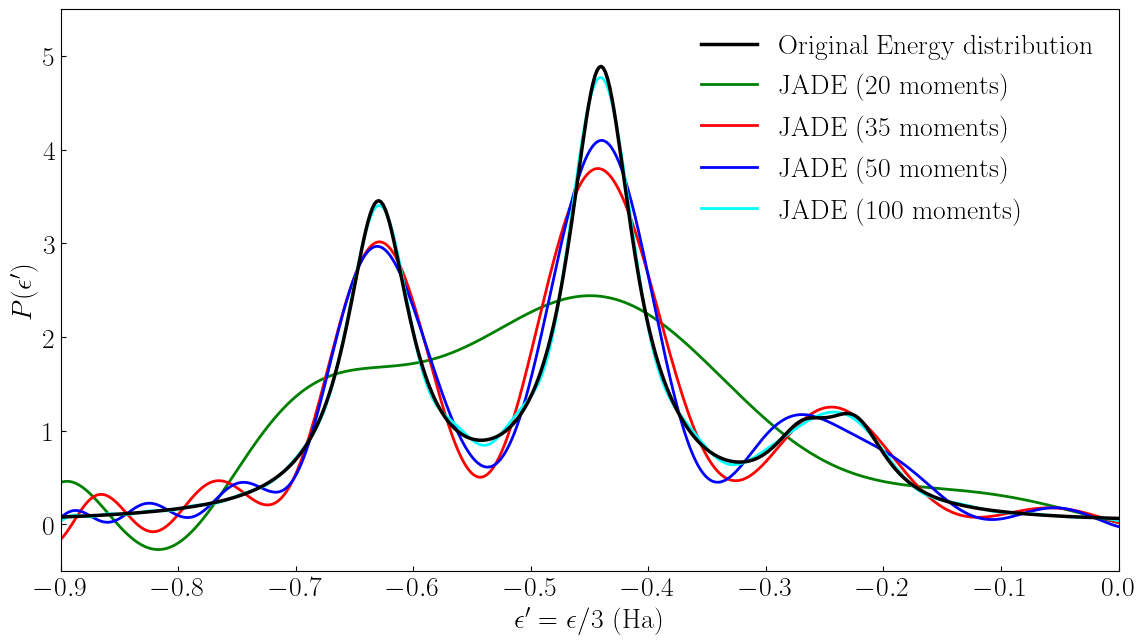}%
      \label{fig:energy_distribution_JADE}%
    }
  \end{minipage}

  \vspace{1ex}

  \caption{
  Estimation of the energy distribution for the Hartree–Fock state of a four-atom hydrogen chain. The $x$-axis is defined as $\epsilon'=\epsilon/3$, reflect the rescaling required by the Chebyshev polynomials, which are defined on $E \in [-1,1]$. Results from the Gram--Charlier expansion and JADE are compared against the exact distribution (black solid line) obtained using the Overlapper software package. For the Gram--Charlier expansion, the number of cumulants was limited to 6 and 12, since higher orders introduce severe oscillations and divergence. Whereas the Gram--Charlier expansion becomes increasingly unstable and inaccurate with the inclusion of additional moments, JADE systematically improves as more moments are included, ultimately converging to the exact distribution. The close agreement highlights JADE’s ability to recover the key features of the energy spectrum from only a finite set of moments.
  }
  \label{fig:energy_distribution}
\end{figure}

Fig.~\ref{fig:energy_distribution} compares the performance of the Gram--Charlier expansion and JADE in estimating the energy distribution of the Hartree–Fock state of a four-atom hydrogen chain. The exact distribution was generated using the software package Overlapper~\cite{44_overlapper}. This distribution serves as our reference, against which the estimating methods are compared. Results from the Gram--Charlier expansion are shown in Fig.~\ref{fig:energy_distribution_GC}, where the number of cumulants was deliberately restricted to 6 and 12, since including higher orders leads to severe oscillations. In contrast, Fig.~\ref{fig:energy_distribution_JADE} presents JADE estimates obtained with an increasing number of moments, ranging from 20 to 100. The comparison highlights a striking divergence: while the Gram--Charlier expansion deteriorates with the inclusion of additional information, JADE systematically improves as more moments are incorporated, ultimately converging to the exact distribution.

An additional and important advantage of JADE is its tunable resolution, controlled by the number of moments used in the estimation. A large number of moments yields a high-resolution reconstruction, while a smaller set provides a coarse but still informative estimate. Even this approximate estimate is sufficient for comparison with a classically computed energy reference, enabling efficient decisions about whether a computationally expensive QPE run is likely to outperform classical methods~\cite{9_Initial}. The Hartree–Fock state in Fig.~\ref{fig:energy_distribution} is a good example; JADE's accurate estimation reveals its multi-peak distribution, confirming that it is a poor initial state because its probability density is not concentrated in the low-energy region. This flexibility allows JADE to serve as an efficient classical pre-check to determine the necessity of a QPE run. In the fault-tolerant quantum computing (FTQC) era, where quantum computational resources will be extremely valuable, JADE is expected to play a key role in conserving them by identifying cases where an expensive QPE run is unlikely to outperform classical methods.

\subsection{Various types of distribution}
\label{sec:various_distribution}

\begin{figure*}[t!]
  \centering
  \begin{minipage}[b]{0.5134\textwidth}
    \subfloat[Bimodal distribution]{%
      \includegraphics[width=\linewidth]{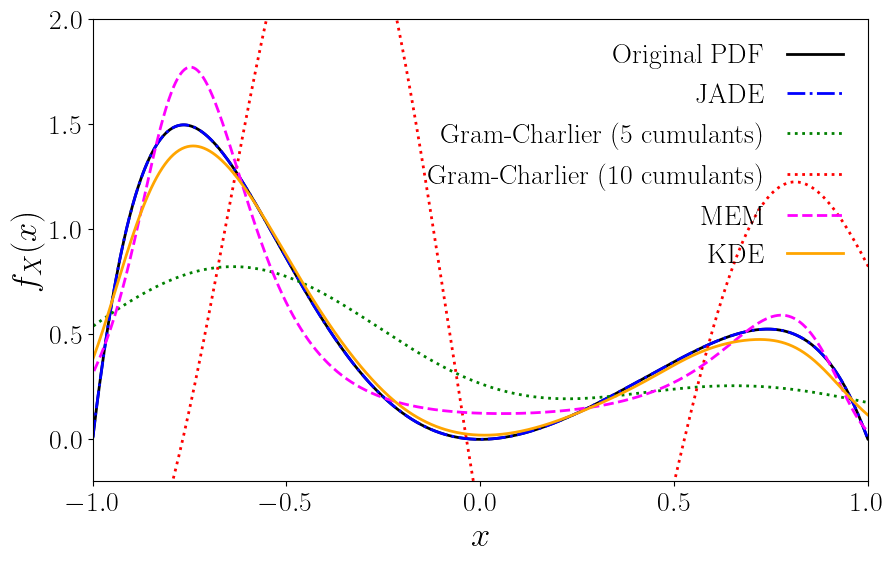}%
      \label{fig:2_plots_a}%
    }
  \end{minipage}\hspace{-0.01\textwidth}
  \begin{minipage}[b]{0.49\textwidth}
    \subfloat[Multimodal distribution]{%
      \includegraphics[width=\linewidth]{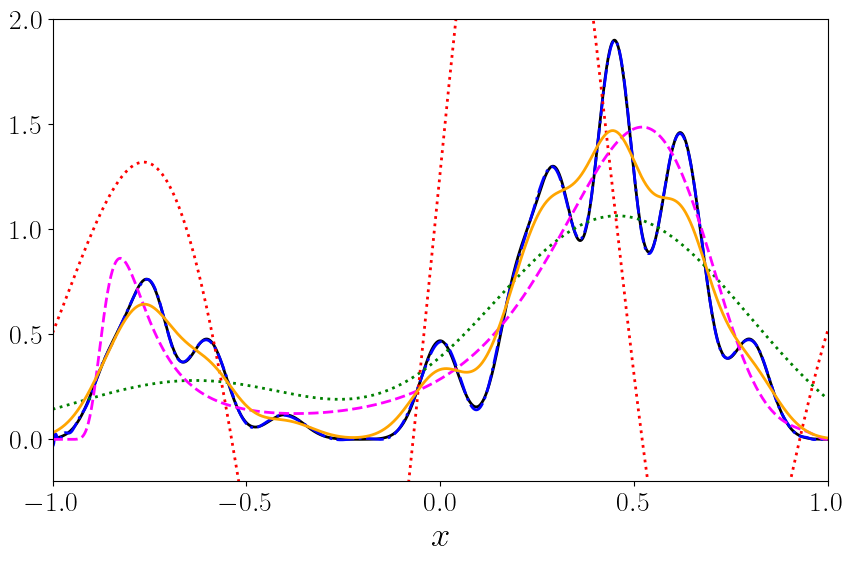}%
      \label{fig:2_plots_b}%
    }
  \end{minipage}

  \vspace{1ex}

  \begin{minipage}[b]{0.5134\textwidth}
    \subfloat[Uniform distribution]{%
      \includegraphics[width=\linewidth]{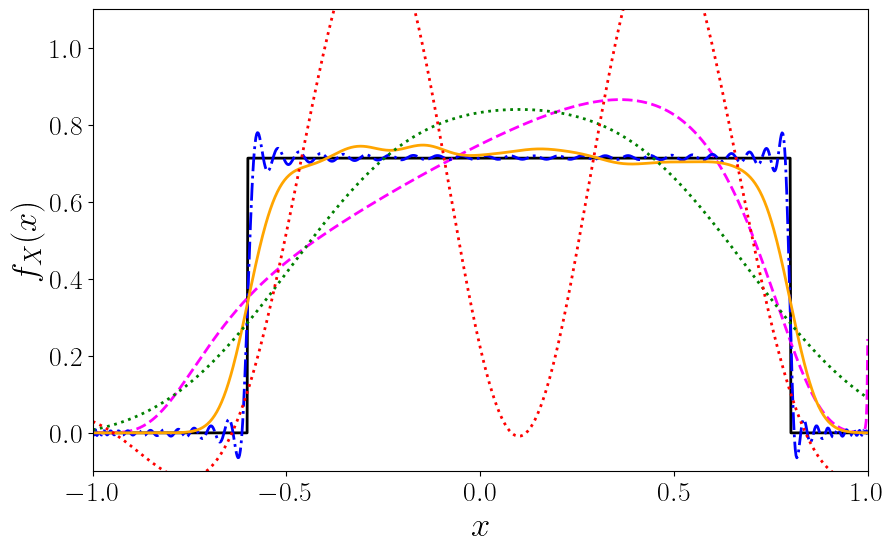}%
      \label{fig:2_plots_c}%
    }
  \end{minipage}\hspace{-0.01\textwidth}
  \begin{minipage}[b]{0.49\textwidth}
    \subfloat[Sigmoid-like function]{%
      \includegraphics[width=\linewidth]{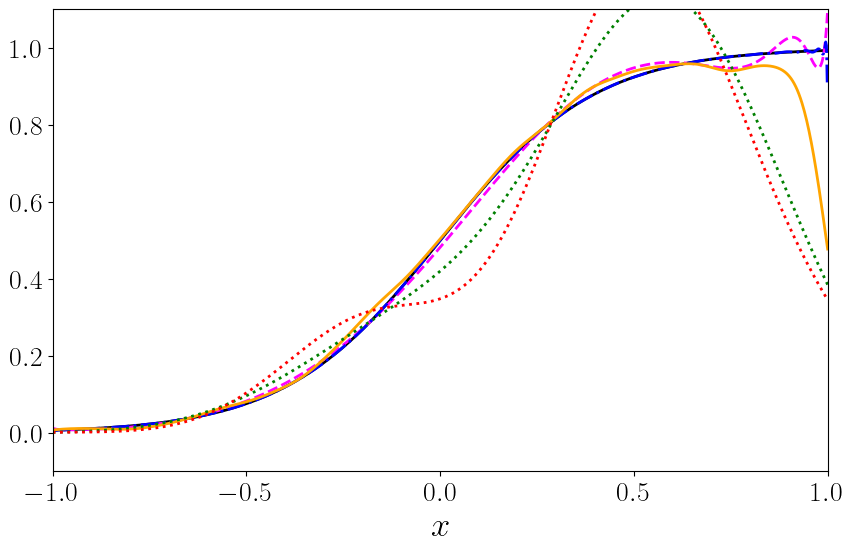}%
      \label{fig:2_plots_d}%
    }
  \end{minipage}

  \caption{
    (a) A bimodal polynomial function, 
    $f_X(x)=-\tfrac{21}{8}(x-1)(x+1)(x^4 - x^3 + x^2)$;
    (b) A randomly generated multimodal distribution using a Gaussian kernel;
    (c) A asymmetric uniform distribution with discontinuities at $x=-0.6$ and $x=0.8$;
    (d) A sigmoid-like function, $f_X(x)=\tfrac{1}{1+\exp(-5x)}$.
    The black solid line shows the original PDF. The blue dashed line denotes the PDF estimated by JADE. The green and red dashed lines correspond to the Gram--Charlier expansion with 5 and 10 cumulants, respectively. The pink dashed line indicates the MEM result, with its number of parameters equal to the number of moments used by JADE, and the orange solid line shows the KDE estimate obtained from 10,000 samples.}
  \label{fig:2plots-wide}
\end{figure*}

In Section~\ref{sec:energy_distribution}, we demonstrated that JADE can accurately estimate the energy distributions of a molecular system and discussed its practical applications. Since JADE is grounded in the Fourier transform relationship between PDFs and their characteristic functions, its accuracy is not limited to energy distributions but extends to general PDFs. In this section, we evaluate JADE’s performance on four complicated distributions that are particularly challenging for conventional approaches. This demonstrates that JADE’s applicability reaches beyond quantum computing into a wide range of domains.

Fig. \ref{fig:2plots-wide} compares four methods (JADE, Gram--Charlier, MEM, and KDE) on four distinct PDFs. Specifically, Fig.~\ref{fig:2_plots_a} shows a polynomial bimodal distribution, Fig.~\ref{fig:2_plots_b} a complex multimodal distribution generated with a Gaussian kernel, Fig.~\ref{fig:2_plots_c} an asymmetric uniform distribution designed to test robustness near singularities, and Fig.~\ref{fig:2_plots_d} a sigmoid-like function commonly used in machine learning.
To ensure a fair comparison, we applied the Gram--Charlier expansion with 5 and 10 cumulants, evaluated MEM under the same moment constraints as JADE, and KDE with 10,000 samples using a Gaussian kernel. These are compared with JADE estimates computed from a finite set of moments.

\textbf{Bimodal distribution (Fig.~\ref{fig:2_plots_a}):}
 This distribution poses difficulties because a simple Gaussian mixture does not reasonably approximate the two modes, and the low-density region between peaks is easily misrepresented. JADE, using 20 moments, reproduces the original distribution with high fidelity, as evidenced by the near overlap of the blue dotted and black solid curves. By contrast, the Gram--Charlier expansion fails to capture the non-Gaussian features, leading to negative probabilities and oscillations~\cite{4_gram_1,9_Initial}. MEM and KDE provide improved estimates but still deviate: MEM underestimates the low-density region, while KDE suffers from boundary bias and peak attenuation when a single global bandwidth is used.

\textbf{Multimodal distribution (Fig.~\ref{fig:2_plots_b}):}
 Estimating multimodal PDFs is widely regarded as difficult~\cite{8_MEM_2}, yet such distributions arise in many engineering applications~\cite{42_mutlimodal_engineering_1,43_mutlimodal_engineering_2}. The original PDF was generated by applying a Gaussian kernel to a PMF. JADE, with 50 moments, achieves a close match to the original distribution. In contrast, the Gram--Charlier expansion deteriorates when moving from 5 to 10 cumulants, introducing oscillations that diverge from the target PDF. MEM and KDE approximate the global shape but fail to accurately capture individual peaks. For KDE, bandwidth selection remains a persistent challenge, even when guided by Silverman’s rule-of-thumb~\cite{46_silverman, 50_silverman2}, and performance remains suboptimal despite using the correct kernel.

\textbf{Asymmetric uniform distribution (Fig.~\ref{fig:2_plots_c}):}
 This distribution, with discontinuous steps, is particularly difficult to approximate. The Gram--Charlier expansion and MEM both fail to estimate it reliably. Furthermore, while the number of MEM's parameters is equal to the number of moments used by JADE, its reliance on nonlinear optimization makes the process computationally demanding as the number of constraints increases. KDE performs better but exhibits inaccuracies at singularities due to the Gaussian kernel, highlighting the difficulty of kernel selection for arbitrary PDFs. JADE, even though showing the Gibbs phenomenon near discontinuities, provides a significantly better approximation. Using 100 moments, JADE captures the discontinuities far more accurately than other methods, which fail even with additional parameters or larger sample sizes.

\textbf{Sigmoid-like distribution (Fig.~\ref{fig:2_plots_d}):}
For smooth distributions such as the normalized sigmoid function \cite{24_sigmoid}, JADE provides a significantly more accurate estimate than the other methods. Using 50 moments, JADE closely matches the original distribution. By contrast, the Gram--Charlier expansion fails to capture the correct curvature and deviates substantially from the target PDF. MEM and KDE perform reasonably well in the central region, but both struggle near the right boundary: MEM develops oscillations, whereas KDE exhibits boundary bias, resulting in a sharp decrease in the estimated density at the boundary.

Taken together, the results in Fig.~\ref{fig:energy_distribution} and Fig.~\ref{fig:2plots-wide} demonstrate that JADE achieves high accuracy in estimations based only on moments, both for the energy distributions of quantum states and for diverse classical distributions. 

\section{Conclusion}
\label{sec:Conclusion}

In this paper, we propose JADE, a novel quantum-inspired method for estimating energy distributions using only the finite set of moments of a  Hamiltonian. Since the energy distribution of a quantum state plays a central role in efficiently estimating ground-state energies in quantum computing, an accurate and practical estimation method is highly valuable. Our numerical results show that JADE can accurately reproduce the energy distribution of a four-atom hydrogen chain using only a finite set of its moments. Beyond quantum systems, we also demonstrated JADE’s superiority over conventional methods on four complex PDFs, underscoring its broad applicability.

By comparing the energy distribution of a quantum state with classically obtained ground-state energies, one can determine whether applying QPE offers an advantage over classical approaches. From this perspective, JADE provides a powerful moment-based framework for efficiently estimating ground-state energies. More generally, its ability to yield accurate estimates for diverse and complicated distributions—where conventional approaches fail—suggests strong potential for applications beyond quantum computing.

The primary contribution of this work is the simultaneous achievement of high accuracy and wide applicability under the constrained condition of using only moments. This distinguishes JADE from other classical approaches: the Gram--Charlier and Edgeworth expansions produce poor estimates under the same constraints, MEM requires expensive parametric nonlinear optimization, and KDE relies on sample data and kernel selection. In contrast, JADE offers a non-parametric, mathematically simple, closed-form expression and is proven to be the optimal solution minimizing weighted $L_2$ loss in function approximation. This highlights its promise as a general-purpose tool for PDF estimation not only in quantum computing but also in modern engineering fields such as machine learning.

Another key advantage of JADE is efficiency, as its performance is directly tied to the cost of computing moments. While computing Hamiltonian moments is already far less demanding than full diagonalization, it can be further accelerated by techniques such as tensor networks and matrix product states (MPS). Recent work also demonstrates that polynomial-approximation methods can compute moments efficiently~\cite{48_park2025quadraticallyshallow}. Combining these techniques with JADE further enhances its practicality.
Looking ahead, several promising research directions emerge. JADE could be applied to estimate energy distributions of quantum states and benchmarked against classically obtained ground-state energies to assess quantum advantage. It could also be directly compared with existing quantum algorithms of similar accuracy to quantify gains in resource efficiency. 

A particularly intriguing direction is moving from evaluating initial state quality to developing methods that actively improve prepared states, thereby enabling accurate energy estimation even from arbitrary states without explicit quantum chemistry calculations. Finally, building on JADE’s demonstrated success across diverse distributions, future research will explore its deeper connections to machine learning. By establishing a formal mathematical link and developing kernel functions inspired by JADE’s analytic structure, we aim to show that its role in machine learning can be not only applicational but also foundational.
$$
$$

\begin{acknowledgments}

We thank Stepan Fomichev for providing the moments data and in-depth discussion. 
This work received partial support from multiple sources:
[1] Basic Science Research Program through the National Research Foundation of Korea (NRF), funded by the Ministry of Science and ICT (RS-2023-NR068116, RS-2023-NR119931, RS-2025-03532992, RS-2025-07882969). [2] Institute for Information \& Communications Technology Promotion (IITP) grant funded by the Korea government (MSIP) (No. 2019-0-00003), which focuses on the research and development of core technologies for programming, running, implementing, and validating fault-tolerant quantum computing systems. [3] Ministry of Trade, Industry, and Energy (MOTIE), Korea, which supported this research under the Industrial Innovation Infrastructure Development Project (Project No. RS-2024-00466693). [4] Korean ARPA-H Project via the Korea Health Industry Development Institute (KHIDI); Ministry of Health and Welfare, Republic of Korea (RS-2025-25456722). [5] Yonsei University Research Fund under project number 2025-22-0140. 
\end{acknowledgments}

\bibliographystyle{apsrev4-2}
\bibliography{ref_JADE}

\begin{thebibliography}{31}%
\makeatletter
\providecommand \@ifxundefined [1]{%
 \@ifx{#1\undefined}
}%
\providecommand \@ifnum [1]{%
 \ifnum #1\expandafter \@firstoftwo
 \else \expandafter \@secondoftwo
 \fi
}%
\providecommand \@ifx [1]{%
 \ifx #1\expandafter \@firstoftwo
 \else \expandafter \@secondoftwo
 \fi
}%
\providecommand \natexlab [1]{#1}%
\providecommand \enquote  [1]{``#1''}%
\providecommand \bibnamefont  [1]{#1}%
\providecommand \bibfnamefont [1]{#1}%
\providecommand \citenamefont [1]{#1}%
\providecommand \href@noop [0]{\@secondoftwo}%
\providecommand \href [0]{\begingroup \@sanitize@url \@href}%
\providecommand \@href[1]{\@@startlink{#1}\@@href}%
\providecommand \@@href[1]{\endgroup#1\@@endlink}%
\providecommand \@sanitize@url [0]{\catcode `\\12\catcode `\$12\catcode `\&12\catcode `\#12\catcode `\^12\catcode `\_12\catcode `\%12\relax}%
\providecommand \@@startlink[1]{}%
\providecommand \@@endlink[0]{}%
\providecommand \url  [0]{\begingroup\@sanitize@url \@url }%
\providecommand \@url [1]{\endgroup\@href {#1}{\urlprefix }}%
\providecommand \urlprefix  [0]{URL }%
\providecommand \Eprint [0]{\href }%
\providecommand \doibase [0]{https://doi.org/}%
\providecommand \selectlanguage [0]{\@gobble}%
\providecommand \bibinfo  [0]{\@secondoftwo}%
\providecommand \bibfield  [0]{\@secondoftwo}%
\providecommand \translation [1]{[#1]}%
\providecommand \BibitemOpen [0]{}%
\providecommand \bibitemStop [0]{}%
\providecommand \bibitemNoStop [0]{.\EOS\space}%
\providecommand \EOS [0]{\spacefactor3000\relax}%
\providecommand \BibitemShut  [1]{\csname bibitem#1\endcsname}%
\let\auto@bib@innerbib\@empty
\bibitem [{\citenamefont {Fomichev}\ \emph {et~al.}(2024{\natexlab{a}})\citenamefont {Fomichev}, \citenamefont {Hejazi}, \citenamefont {Zini}, \citenamefont {Kiser}, \citenamefont {Fraxanet}, \citenamefont {Casares}, \citenamefont {Delgado}, \citenamefont {Huh}, \citenamefont {Voigt}, \citenamefont {Mueller},\ and\ \citenamefont {Arrazola}}]{9_Initial}%
  \BibitemOpen
  \bibfield  {author} {\bibinfo {author} {\bibfnamefont {S.}~\bibnamefont {Fomichev}}, \bibinfo {author} {\bibfnamefont {K.}~\bibnamefont {Hejazi}}, \bibinfo {author} {\bibfnamefont {M.~S.}\ \bibnamefont {Zini}}, \bibinfo {author} {\bibfnamefont {M.}~\bibnamefont {Kiser}}, \bibinfo {author} {\bibfnamefont {J.}~\bibnamefont {Fraxanet}}, \bibinfo {author} {\bibfnamefont {P.~A.~M.}\ \bibnamefont {Casares}}, \bibinfo {author} {\bibfnamefont {A.}~\bibnamefont {Delgado}}, \bibinfo {author} {\bibfnamefont {J.}~\bibnamefont {Huh}}, \bibinfo {author} {\bibfnamefont {A.-C.}\ \bibnamefont {Voigt}}, \bibinfo {author} {\bibfnamefont {J.~E.}\ \bibnamefont {Mueller}},\ and\ \bibinfo {author} {\bibfnamefont {J.~M.}\ \bibnamefont {Arrazola}},\ }\href {https://doi.org/10.1103/PRXQuantum.5.040339} {\bibfield  {journal} {\bibinfo  {journal} {PRX Quantum}\ }\textbf {\bibinfo {volume} {5}},\ \bibinfo {pages} {040339} (\bibinfo {year} {2024}{\natexlab{a}})}\BibitemShut {NoStop}%
\bibitem [{\citenamefont {Berry}\ \emph {et~al.}(2025)\citenamefont {Berry}, \citenamefont {Tong}, \citenamefont {Khattar}, \citenamefont {White}, \citenamefont {Kim}, \citenamefont {Low}, \citenamefont {Boixo}, \citenamefont {Ding}, \citenamefont {Lin}, \citenamefont {Lee}, \citenamefont {Chan}, \citenamefont {Babbush},\ and\ \citenamefont {Rubin}}]{45_initial_prep_important}%
  \BibitemOpen
  \bibfield  {author} {\bibinfo {author} {\bibfnamefont {D.~W.}\ \bibnamefont {Berry}}, \bibinfo {author} {\bibfnamefont {Y.}~\bibnamefont {Tong}}, \bibinfo {author} {\bibfnamefont {T.}~\bibnamefont {Khattar}}, \bibinfo {author} {\bibfnamefont {A.}~\bibnamefont {White}}, \bibinfo {author} {\bibfnamefont {T.~I.}\ \bibnamefont {Kim}}, \bibinfo {author} {\bibfnamefont {G.~H.}\ \bibnamefont {Low}}, \bibinfo {author} {\bibfnamefont {S.}~\bibnamefont {Boixo}}, \bibinfo {author} {\bibfnamefont {Z.}~\bibnamefont {Ding}}, \bibinfo {author} {\bibfnamefont {L.}~\bibnamefont {Lin}}, \bibinfo {author} {\bibfnamefont {S.}~\bibnamefont {Lee}}, \bibinfo {author} {\bibfnamefont {G.~K.-L.}\ \bibnamefont {Chan}}, \bibinfo {author} {\bibfnamefont {R.}~\bibnamefont {Babbush}},\ and\ \bibinfo {author} {\bibfnamefont {N.~C.}\ \bibnamefont {Rubin}},\ }\href {https://doi.org/10.1103/PRXQuantum.6.020327} {\bibfield  {journal} {\bibinfo  {journal} {PRX Quantum}\ }\textbf {\bibinfo {volume} {6}},\ \bibinfo {pages} {020327} (\bibinfo
  {year} {2025})}\BibitemShut {NoStop}%
\bibitem [{\citenamefont {Feniou}\ \emph {et~al.}(2024)\citenamefont {Feniou}, \citenamefont {Adjoua}, \citenamefont {Claudon}, \citenamefont {Zylberman}, \citenamefont {Giner},\ and\ \citenamefont {Piquemal}}]{51_VQE&QPE}%
  \BibitemOpen
  \bibfield  {author} {\bibinfo {author} {\bibfnamefont {C.}~\bibnamefont {Feniou}}, \bibinfo {author} {\bibfnamefont {O.}~\bibnamefont {Adjoua}}, \bibinfo {author} {\bibfnamefont {B.}~\bibnamefont {Claudon}}, \bibinfo {author} {\bibfnamefont {J.}~\bibnamefont {Zylberman}}, \bibinfo {author} {\bibfnamefont {E.}~\bibnamefont {Giner}},\ and\ \bibinfo {author} {\bibfnamefont {J.-P.}\ \bibnamefont {Piquemal}},\ }\href {https://doi.org/10.1021/acs.jpclett.3c03159} {\bibfield  {journal} {\bibinfo  {journal} {J. Phys. Chem. Lett.}\ }\textbf {\bibinfo {volume} {15}},\ \bibinfo {pages} {3197} (\bibinfo {year} {2024})}\BibitemShut {NoStop}%
\bibitem [{\citenamefont {Grant}\ \emph {et~al.}(2019)\citenamefont {Grant}, \citenamefont {Wossnig}, \citenamefont {Ostaszewski},\ and\ \citenamefont {Benedetti}}]{55_VQE}%
  \BibitemOpen
  \bibfield  {author} {\bibinfo {author} {\bibfnamefont {E.}~\bibnamefont {Grant}}, \bibinfo {author} {\bibfnamefont {L.}~\bibnamefont {Wossnig}}, \bibinfo {author} {\bibfnamefont {M.}~\bibnamefont {Ostaszewski}},\ and\ \bibinfo {author} {\bibfnamefont {M.}~\bibnamefont {Benedetti}},\ }\href {https://doi.org/10.22331/q-2019-12-09-214} {\bibfield  {journal} {\bibinfo  {journal} {Quantum}\ }\textbf {\bibinfo {volume} {3}},\ \bibinfo {pages} {214} (\bibinfo {year} {2019})}\BibitemShut {NoStop}%
\bibitem [{\citenamefont {Baker}\ \emph {et~al.}(2025)\citenamefont {Baker}, \citenamefont {Saxena},\ and\ \citenamefont {Kyaw}}]{52_QPE}%
  \BibitemOpen
  \bibfield  {author} {\bibinfo {author} {\bibfnamefont {J.~S.}\ \bibnamefont {Baker}}, \bibinfo {author} {\bibfnamefont {G.}~\bibnamefont {Saxena}},\ and\ \bibinfo {author} {\bibfnamefont {T.~H.}\ \bibnamefont {Kyaw}},\ }\href@noop {} {\bibinfo {title} {Universal initial state preparation for first quantized quantum simulations}} (\bibinfo {year} {2025}),\ \Eprint {https://arxiv.org/abs/2510.07278} {arXiv:2510.07278} \BibitemShut {NoStop}%
\bibitem [{\citenamefont {Abrams}\ and\ \citenamefont {Lloyd}(1999)}]{56_QPE}%
  \BibitemOpen
  \bibfield  {author} {\bibinfo {author} {\bibfnamefont {D.~S.}\ \bibnamefont {Abrams}}\ and\ \bibinfo {author} {\bibfnamefont {S.}~\bibnamefont {Lloyd}},\ }\href {https://doi.org/10.1103/PhysRevLett.83.5162} {\bibfield  {journal} {\bibinfo  {journal} {Phys. Rev. Lett.}\ }\textbf {\bibinfo {volume} {83}},\ \bibinfo {pages} {5162} (\bibinfo {year} {1999})}\BibitemShut {NoStop}%
\bibitem [{\citenamefont {Ollitrault}\ \emph {et~al.}(2024)\citenamefont {Ollitrault}, \citenamefont {Cortes}, \citenamefont {Gonthier}, \citenamefont {Parrish}, \citenamefont {Rocca}, \citenamefont {Anselmetti}, \citenamefont {Degroote}, \citenamefont {Moll}, \citenamefont {Santagati},\ and\ \citenamefont {Streif}}]{57_QPE}%
  \BibitemOpen
  \bibfield  {author} {\bibinfo {author} {\bibfnamefont {P.~J.}\ \bibnamefont {Ollitrault}}, \bibinfo {author} {\bibfnamefont {C.~L.}\ \bibnamefont {Cortes}}, \bibinfo {author} {\bibfnamefont {J.~F.}\ \bibnamefont {Gonthier}}, \bibinfo {author} {\bibfnamefont {R.~M.}\ \bibnamefont {Parrish}}, \bibinfo {author} {\bibfnamefont {D.}~\bibnamefont {Rocca}}, \bibinfo {author} {\bibfnamefont {G.}~\bibnamefont {Anselmetti}}, \bibinfo {author} {\bibfnamefont {M.}~\bibnamefont {Degroote}}, \bibinfo {author} {\bibfnamefont {N.}~\bibnamefont {Moll}}, \bibinfo {author} {\bibfnamefont {R.}~\bibnamefont {Santagati}},\ and\ \bibinfo {author} {\bibfnamefont {M.}~\bibnamefont {Streif}},\ }\href {https://doi.org/10.1103/PhysRevLett.133.250601} {\bibfield  {journal} {\bibinfo  {journal} {Phys. Rev. Lett.}\ }\textbf {\bibinfo {volume} {133}},\ \bibinfo {pages} {250601} (\bibinfo {year} {2024})}\BibitemShut {NoStop}%
\bibitem [{\citenamefont {Yu}\ \emph {et~al.}(2025)\citenamefont {Yu}, \citenamefont {{Robledo Moreno}}, \citenamefont {Iosue}, \citenamefont {Bertels}, \citenamefont {Claudino}, \citenamefont {Fuller}, \citenamefont {Groszkowski}, \citenamefont {Humble}, \citenamefont {Jurcevic}, \citenamefont {Kirby}, \citenamefont {Maier}, \citenamefont {Motta}, \citenamefont {Pokharel}, \citenamefont {Seif}, \citenamefont {Shehata}, \citenamefont {Sung}, \citenamefont {Tran}, \citenamefont {Tripathi}, \citenamefont {Mezzacapo},\ and\ \citenamefont {Sharma}}]{53_KQD}%
  \BibitemOpen
  \bibfield  {author} {\bibinfo {author} {\bibfnamefont {J.}~\bibnamefont {Yu}}, \bibinfo {author} {\bibfnamefont {J.}~\bibnamefont {{Robledo Moreno}}}, \bibinfo {author} {\bibfnamefont {J.~T.}\ \bibnamefont {Iosue}}, \bibinfo {author} {\bibfnamefont {L.}~\bibnamefont {Bertels}}, \bibinfo {author} {\bibfnamefont {D.}~\bibnamefont {Claudino}}, \bibinfo {author} {\bibfnamefont {B.}~\bibnamefont {Fuller}}, \bibinfo {author} {\bibfnamefont {P.}~\bibnamefont {Groszkowski}}, \bibinfo {author} {\bibfnamefont {T.~S.}\ \bibnamefont {Humble}}, \bibinfo {author} {\bibfnamefont {P.}~\bibnamefont {Jurcevic}}, \bibinfo {author} {\bibfnamefont {W.}~\bibnamefont {Kirby}}, \bibinfo {author} {\bibfnamefont {T.~A.}\ \bibnamefont {Maier}}, \bibinfo {author} {\bibfnamefont {M.}~\bibnamefont {Motta}}, \bibinfo {author} {\bibfnamefont {B.}~\bibnamefont {Pokharel}}, \bibinfo {author} {\bibfnamefont {A.}~\bibnamefont {Seif}}, \bibinfo {author} {\bibfnamefont {A.}~\bibnamefont {Shehata}}, \bibinfo {author} {\bibfnamefont {K.~J.}\
  \bibnamefont {Sung}}, \bibinfo {author} {\bibfnamefont {M.~C.}\ \bibnamefont {Tran}}, \bibinfo {author} {\bibfnamefont {V.}~\bibnamefont {Tripathi}}, \bibinfo {author} {\bibfnamefont {A.}~\bibnamefont {Mezzacapo}},\ and\ \bibinfo {author} {\bibfnamefont {K.}~\bibnamefont {Sharma}},\ }\href@noop {} {\bibinfo {title} {Quantum-centric algorithm for sample-based {Krylov} diagonalization}} (\bibinfo {year} {2025}),\ \Eprint {https://arxiv.org/abs/2501.09702} {arXiv:2501.09702} \BibitemShut {NoStop}%
\bibitem [{\citenamefont {Kirby}\ \emph {et~al.}(2023)\citenamefont {Kirby}, \citenamefont {Motta},\ and\ \citenamefont {Mezzacapo}}]{58_KQD}%
  \BibitemOpen
  \bibfield  {author} {\bibinfo {author} {\bibfnamefont {W.}~\bibnamefont {Kirby}}, \bibinfo {author} {\bibfnamefont {M.}~\bibnamefont {Motta}},\ and\ \bibinfo {author} {\bibfnamefont {A.}~\bibnamefont {Mezzacapo}},\ }\href {https://doi.org/10.22331/q-2023-05-23-1018} {\bibfield  {journal} {\bibinfo  {journal} {Quantum}\ }\textbf {\bibinfo {volume} {7}},\ \bibinfo {pages} {1018} (\bibinfo {year} {2023})}\BibitemShut {NoStop}%
\bibitem [{\citenamefont {Berberan-Santos}(2007)}]{11_gram_2}%
  \BibitemOpen
  \bibfield  {author} {\bibinfo {author} {\bibfnamefont {M.~N.}\ \bibnamefont {Berberan-Santos}},\ }\href {https://doi.org/10.1007/s10910-006-9134-5} {\bibfield  {journal} {\bibinfo  {journal} {J. Math. Chem.}\ }\textbf {\bibinfo {volume} {42}},\ \bibinfo {pages} {585} (\bibinfo {year} {2007})}\BibitemShut {NoStop}%
\bibitem [{\citenamefont {Blinnikov}\ and\ \citenamefont {Moessner}(1998)}]{blinnikov1998expansions}%
  \BibitemOpen
  \bibfield  {author} {\bibinfo {author} {\bibfnamefont {S.}~\bibnamefont {Blinnikov}}\ and\ \bibinfo {author} {\bibfnamefont {R.}~\bibnamefont {Moessner}},\ }\href {https://doi.org/10.1051/aas:1998221} {\bibfield  {journal} {\bibinfo  {journal} {Astron. Astrophys., Suppl. Ser.}\ }\textbf {\bibinfo {volume} {130}},\ \bibinfo {pages} {193} (\bibinfo {year} {1998})}\BibitemShut {NoStop}%
\bibitem [{\citenamefont {Wang}\ \emph {et~al.}(2022)\citenamefont {Wang}, \citenamefont {Seta},\ and\ \citenamefont {Ikegaya}}]{4_gram_1}%
  \BibitemOpen
  \bibfield  {author} {\bibinfo {author} {\bibfnamefont {W.}~\bibnamefont {Wang}}, \bibinfo {author} {\bibfnamefont {K.}~\bibnamefont {Seta}},\ and\ \bibinfo {author} {\bibfnamefont {N.}~\bibnamefont {Ikegaya}},\ }\href {https://doi.org/10.1016/j.jweia.2022.105227} {\bibfield  {journal} {\bibinfo  {journal} {J. Wind Eng. Ind. Aerodyn.}\ }\textbf {\bibinfo {volume} {231}},\ \bibinfo {pages} {105227} (\bibinfo {year} {2022})}\BibitemShut {NoStop}%
\bibitem [{\citenamefont {Jonsson}\ and\ \citenamefont {Felsberg}(2005)}]{7_MEM_1}%
  \BibitemOpen
  \bibfield  {author} {\bibinfo {author} {\bibfnamefont {E.}~\bibnamefont {Jonsson}}\ and\ \bibinfo {author} {\bibfnamefont {M.}~\bibnamefont {Felsberg}},\ }in\ \href {https://doi.org/10.1007/11499145_50} {\emph {\bibinfo {booktitle} {Image Analysis}}},\ \bibinfo {series} {Lecture Notes in Computer Science}, Vol.\ \bibinfo {volume} {3540},\ \bibinfo {editor} {edited by\ \bibinfo {editor} {\bibfnamefont {H.}~\bibnamefont {Kalviainen}}, \bibinfo {editor} {\bibfnamefont {J.}~\bibnamefont {Parkkinen}},\ and\ \bibinfo {editor} {\bibfnamefont {A.}~\bibnamefont {Kaarna}}}\ (\bibinfo  {publisher} {Springer},\ \bibinfo {address} {Berlin, Heidelberg},\ \bibinfo {year} {2005})\ pp.\ \bibinfo {pages} {491--500}\BibitemShut {NoStop}%
\bibitem [{\citenamefont {Li}\ \emph {et~al.}(2022)\citenamefont {Li}, \citenamefont {Wang}, \citenamefont {Zeng},\ and\ \citenamefont {He}}]{8_MEM_2}%
  \BibitemOpen
  \bibfield  {author} {\bibinfo {author} {\bibfnamefont {G.}~\bibnamefont {Li}}, \bibinfo {author} {\bibfnamefont {Y.~X.}\ \bibnamefont {Wang}}, \bibinfo {author} {\bibfnamefont {Y.}~\bibnamefont {Zeng}},\ and\ \bibinfo {author} {\bibfnamefont {W.~X.}\ \bibnamefont {He}},\ }\href {https://doi.org/10.1016/j.apm.2021.09.029} {\bibfield  {journal} {\bibinfo  {journal} {Appl. Math. Model.}\ }\textbf {\bibinfo {volume} {102}},\ \bibinfo {pages} {137} (\bibinfo {year} {2022})}\BibitemShut {NoStop}%
\bibitem [{\citenamefont {Jaynes}(1957{\natexlab{a}})}]{14_MEM_unbias}%
  \BibitemOpen
  \bibfield  {author} {\bibinfo {author} {\bibfnamefont {E.~T.}\ \bibnamefont {Jaynes}},\ }\href {https://doi.org/10.1103/PhysRev.106.620} {\bibfield  {journal} {\bibinfo  {journal} {Phys. Rev.}\ }\textbf {\bibinfo {volume} {106}},\ \bibinfo {pages} {620} (\bibinfo {year} {1957}{\natexlab{a}})}\BibitemShut {NoStop}%
\bibitem [{\citenamefont {Jaynes}(1957{\natexlab{b}})}]{54_MEM_unbias_2}%
  \BibitemOpen
  \bibfield  {author} {\bibinfo {author} {\bibfnamefont {E.~T.}\ \bibnamefont {Jaynes}},\ }\href {https://doi.org/10.1103/PhysRev.108.171} {\bibfield  {journal} {\bibinfo  {journal} {Phys. Rev.}\ }\textbf {\bibinfo {volume} {108}},\ \bibinfo {pages} {171} (\bibinfo {year} {1957}{\natexlab{b}})}\BibitemShut {NoStop}%
\bibitem [{\citenamefont {Chen}(2017)}]{5_KDE_1}%
  \BibitemOpen
  \bibfield  {author} {\bibinfo {author} {\bibfnamefont {Y.-C.}\ \bibnamefont {Chen}},\ }\href {https://doi.org/10.1080/24709360.2017.1396742} {\bibfield  {journal} {\bibinfo  {journal} {Biostat. Epidemiol.}\ }\textbf {\bibinfo {volume} {1}},\ \bibinfo {pages} {161} (\bibinfo {year} {2017})}\BibitemShut {NoStop}%
\bibitem [{\citenamefont {W{\k{e}}glarczyk}(2018)}]{6_KDE_2}%
  \BibitemOpen
  \bibfield  {author} {\bibinfo {author} {\bibfnamefont {S.}~\bibnamefont {W{\k{e}}glarczyk}},\ }\href {https://doi.org/10.1051/itmconf/20182300037} {\bibfield  {journal} {\bibinfo  {journal} {ITM Web Conf.}\ }\textbf {\bibinfo {volume} {23}},\ \bibinfo {pages} {00037} (\bibinfo {year} {2018})}\BibitemShut {NoStop}%
\bibitem [{\citenamefont {Chen}\ \emph {et~al.}(2024)\citenamefont {Chen}, \citenamefont {He}, \citenamefont {Cheng}, \citenamefont {Fournier-Viger},\ and\ \citenamefont {Huang}}]{16_KDE_3}%
  \BibitemOpen
  \bibfield  {author} {\bibinfo {author} {\bibfnamefont {J.-Q.}\ \bibnamefont {Chen}}, \bibinfo {author} {\bibfnamefont {Y.-L.}\ \bibnamefont {He}}, \bibinfo {author} {\bibfnamefont {Y.-C.}\ \bibnamefont {Cheng}}, \bibinfo {author} {\bibfnamefont {P.}~\bibnamefont {Fournier-Viger}},\ and\ \bibinfo {author} {\bibfnamefont {J.~Z.}\ \bibnamefont {Huang}},\ }\href {https://doi.org/10.1016/j.engappai.2024.107979} {\bibfield  {journal} {\bibinfo  {journal} {Eng. Appl. Artif. Intell.}\ }\textbf {\bibinfo {volume} {132}},\ \bibinfo {pages} {107979} (\bibinfo {year} {2024})}\BibitemShut {NoStop}%
\bibitem [{\citenamefont {Manolakis}\ \emph {et~al.}(2005)\citenamefont {Manolakis}, \citenamefont {Ingle},\ and\ \citenamefont {Kogon}}]{12_fourier}%
  \BibitemOpen
  \bibfield  {author} {\bibinfo {author} {\bibfnamefont {D.~G.}\ \bibnamefont {Manolakis}}, \bibinfo {author} {\bibfnamefont {V.~K.}\ \bibnamefont {Ingle}},\ and\ \bibinfo {author} {\bibfnamefont {S.~M.}\ \bibnamefont {Kogon}},\ }\href@noop {} {\emph {\bibinfo {title} {Statistical and Adaptive Signal Processing}}}\ (\bibinfo  {publisher} {Artech House},\ \bibinfo {address} {Norwood, MA},\ \bibinfo {year} {2005})\BibitemShut {NoStop}%
\bibitem [{\citenamefont {Davies}(1973)}]{36_characteristic}%
  \BibitemOpen
  \bibfield  {author} {\bibinfo {author} {\bibfnamefont {R.~B.}\ \bibnamefont {Davies}},\ }\href {https://doi.org/10.1093/biomet/60.2.415} {\bibfield  {journal} {\bibinfo  {journal} {Biometrika}\ }\textbf {\bibinfo {volume} {60}},\ \bibinfo {pages} {415} (\bibinfo {year} {1973})}\BibitemShut {NoStop}%
\bibitem [{\citenamefont {Cuyt}\ \emph {et~al.}(2008)\citenamefont {Cuyt}, \citenamefont {Petersen}, \citenamefont {Verdonk}, \citenamefont {Waadeland},\ and\ \citenamefont {Jones}}]{20_Jacobi-anger_expansion}%
  \BibitemOpen
  \bibfield  {author} {\bibinfo {author} {\bibfnamefont {A.~A.~M.}\ \bibnamefont {Cuyt}}, \bibinfo {author} {\bibfnamefont {V.~B.}\ \bibnamefont {Petersen}}, \bibinfo {author} {\bibfnamefont {B.}~\bibnamefont {Verdonk}}, \bibinfo {author} {\bibfnamefont {H.}~\bibnamefont {Waadeland}},\ and\ \bibinfo {author} {\bibfnamefont {W.~B.}\ \bibnamefont {Jones}},\ }\href {https://doi.org/10.1007/978-1-4020-6949-9} {\emph {\bibinfo {title} {Handbook of Continued Fractions for Special Functions}}}\ (\bibinfo  {publisher} {Springer},\ \bibinfo {address} {Dordrecht},\ \bibinfo {year} {2008})\BibitemShut {NoStop}%
\bibitem [{\citenamefont {Abramowitz}\ and\ \citenamefont {Stegun}(1965)}]{32_bessel}%
  \BibitemOpen
  \bibinfo {editor} {\bibfnamefont {M.}~\bibnamefont {Abramowitz}}\ and\ \bibinfo {editor} {\bibfnamefont {I.~A.}\ \bibnamefont {Stegun}},\ eds.,\ \href@noop {} {\emph {\bibinfo {title} {Handbook of Mathematical Functions: With Formulas, Graphs, and Mathematical Tables}}}\ (\bibinfo  {publisher} {Dover},\ \bibinfo {address} {New York},\ \bibinfo {year} {1965})\BibitemShut {NoStop}%
\bibitem [{\citenamefont {Mason}\ and\ \citenamefont {Handscomb}(2002)}]{19_chev}%
  \BibitemOpen
  \bibfield  {author} {\bibinfo {author} {\bibfnamefont {J.~C.}\ \bibnamefont {Mason}}\ and\ \bibinfo {author} {\bibfnamefont {D.~C.}\ \bibnamefont {Handscomb}},\ }\href {https://doi.org/10.1201/9781420036114} {\emph {\bibinfo {title} {Chebyshev Polynomials}}}\ (\bibinfo  {publisher} {Chapman and Hall/CRC},\ \bibinfo {address} {Boca Raton},\ \bibinfo {year} {2002})\BibitemShut {NoStop}%
\bibitem [{\citenamefont {Fomichev}\ \emph {et~al.}(2024{\natexlab{b}})\citenamefont {Fomichev}, \citenamefont {Hejazi}, \citenamefont {Fraxanet},\ and\ \citenamefont {Arrazola}}]{44_overlapper}%
  \BibitemOpen
  \bibfield  {author} {\bibinfo {author} {\bibfnamefont {S.}~\bibnamefont {Fomichev}}, \bibinfo {author} {\bibfnamefont {K.}~\bibnamefont {Hejazi}}, \bibinfo {author} {\bibfnamefont {J.}~\bibnamefont {Fraxanet}},\ and\ \bibinfo {author} {\bibfnamefont {J.~M.}\ \bibnamefont {Arrazola}},\ }\href@noop {} {\bibinfo {title} {Overlapper}},\ \bibinfo {howpublished} {\url{https://github.com/XanaduAI/Overlapper/}} (\bibinfo {year} {2024}{\natexlab{b}}),\ \bibinfo {note} {{GitHub} repository}\BibitemShut {NoStop}%
\bibitem [{\citenamefont {Haider}\ \emph {et~al.}(2009)\citenamefont {Haider}, \citenamefont {Harichandran},\ and\ \citenamefont {Dwaikat}}]{42_mutlimodal_engineering_1}%
  \BibitemOpen
  \bibfield  {author} {\bibinfo {author} {\bibfnamefont {S.~W.}\ \bibnamefont {Haider}}, \bibinfo {author} {\bibfnamefont {R.~S.}\ \bibnamefont {Harichandran}},\ and\ \bibinfo {author} {\bibfnamefont {M.~B.}\ \bibnamefont {Dwaikat}},\ }\href {https://doi.org/10.1061/(ASCE)TE.1943-5436.0000077} {\bibfield  {journal} {\bibinfo  {journal} {J. Transp. Eng.}\ }\textbf {\bibinfo {volume} {135}},\ \bibinfo {pages} {974} (\bibinfo {year} {2009})}\BibitemShut {NoStop}%
\bibitem [{\citenamefont {He}\ \emph {et~al.}(2016)\citenamefont {He}, \citenamefont {Guan},\ and\ \citenamefont {Jha}}]{43_mutlimodal_engineering_2}%
  \BibitemOpen
  \bibfield  {author} {\bibinfo {author} {\bibfnamefont {J.}~\bibnamefont {He}}, \bibinfo {author} {\bibfnamefont {X.}~\bibnamefont {Guan}},\ and\ \bibinfo {author} {\bibfnamefont {R.}~\bibnamefont {Jha}},\ }\href {https://doi.org/10.1109/TR.2016.2604121} {\bibfield  {journal} {\bibinfo  {journal} {IEEE Trans. Reliab.}\ }\textbf {\bibinfo {volume} {65}},\ \bibinfo {pages} {1724} (\bibinfo {year} {2016})}\BibitemShut {NoStop}%
\bibitem [{\citenamefont {Silverman}(1986)}]{46_silverman}%
  \BibitemOpen
  \bibfield  {author} {\bibinfo {author} {\bibfnamefont {B.~W.}\ \bibnamefont {Silverman}},\ }\href@noop {} {\emph {\bibinfo {title} {Density Estimation for Statistics and Data Analysis}}}\ (\bibinfo  {publisher} {Chapman and Hall},\ \bibinfo {address} {London},\ \bibinfo {year} {1986})\BibitemShut {NoStop}%
\bibitem [{\citenamefont {Harpole}\ \emph {et~al.}(2014)\citenamefont {Harpole}, \citenamefont {Woods}, \citenamefont {Rodebaugh}, \citenamefont {Levinson},\ and\ \citenamefont {Lenze}}]{50_silverman2}%
  \BibitemOpen
  \bibfield  {author} {\bibinfo {author} {\bibfnamefont {J.~K.}\ \bibnamefont {Harpole}}, \bibinfo {author} {\bibfnamefont {C.~M.}\ \bibnamefont {Woods}}, \bibinfo {author} {\bibfnamefont {T.~L.}\ \bibnamefont {Rodebaugh}}, \bibinfo {author} {\bibfnamefont {C.~A.}\ \bibnamefont {Levinson}},\ and\ \bibinfo {author} {\bibfnamefont {E.~J.}\ \bibnamefont {Lenze}},\ }\href {https://doi.org/10.1037/a0036850} {\bibfield  {journal} {\bibinfo  {journal} {Psychol. Methods}\ }\textbf {\bibinfo {volume} {19}},\ \bibinfo {pages} {428} (\bibinfo {year} {2014})}\BibitemShut {NoStop}%
\bibitem [{\citenamefont {Glorot}\ \emph {et~al.}(2011)\citenamefont {Glorot}, \citenamefont {Bordes},\ and\ \citenamefont {Bengio}}]{24_sigmoid}%
  \BibitemOpen
  \bibfield  {author} {\bibinfo {author} {\bibfnamefont {X.}~\bibnamefont {Glorot}}, \bibinfo {author} {\bibfnamefont {A.}~\bibnamefont {Bordes}},\ and\ \bibinfo {author} {\bibfnamefont {Y.}~\bibnamefont {Bengio}},\ }in\ \href {https://proceedings.mlr.press/v15/glorot11a.html} {\emph {\bibinfo {booktitle} {Proceedings of the Fourteenth International Conference on Artificial Intelligence and Statistics}}},\ \bibinfo {series} {JMLR Workshop and Conference Proceedings}, Vol.~\bibinfo {volume} {15},\ \bibinfo {editor} {edited by\ \bibinfo {editor} {\bibfnamefont {G.}~\bibnamefont {Gordon}}, \bibinfo {editor} {\bibfnamefont {D.}~\bibnamefont {Dunson}},\ and\ \bibinfo {editor} {\bibfnamefont {M.}~\bibnamefont {Dudík}}}\ (\bibinfo  {publisher} {PMLR},\ \bibinfo {address} {Fort Lauderdale, FL, USA},\ \bibinfo {year} {2011})\ pp.\ \bibinfo {pages} {315--323}\BibitemShut {NoStop}%
\bibitem [{\citenamefont {Park}\ \emph {et~al.}(2025)\citenamefont {Park}, \citenamefont {Kang}, \citenamefont {Park},\ and\ \citenamefont {Huh}}]{48_park2025quadraticallyshallow}%
  \BibitemOpen
  \bibfield  {author} {\bibinfo {author} {\bibfnamefont {Y.}~\bibnamefont {Park}}, \bibinfo {author} {\bibfnamefont {M.}~\bibnamefont {Kang}}, \bibinfo {author} {\bibfnamefont {C.-Y.}\ \bibnamefont {Park}},\ and\ \bibinfo {author} {\bibfnamefont {J.}~\bibnamefont {Huh}},\ }\href@noop {} {\bibinfo {title} {Quadratically shallow quantum circuits for {Hamiltonian} functions}} (\bibinfo {year} {2025}),\ \Eprint {https://arxiv.org/abs/2510.04059} {arXiv:2510.04059} \BibitemShut {NoStop}%
\end{thebibliography}%

\section*{Appendix}
\appendix
\section{Detailed Procedure}
\label{sec:Ap_A}
In Section~\ref{sec:Methods}, we introduced the JADE for estimating energy distributions. This section provides a more detailed description of that procedure. Since JADE can be applied not only to energy distributions but also to various other distributions, in this section, we present a generalized formulation of JADE for the PDF.

\subsection{Approximating characteristic function}
\label{sec:Ap_A_1}
The $n$th moment of the random variable X is defined as follows:
\begin{equation}
\label{eq:1_raw_moment}
\mu'_n \coloneqq \int_{-\infty}^{\infty}{x^n f_X(x)\, \mathrm{d}x},
\end{equation}
where $n$ is a non-negative integer and $f_X(x)$ is the PDF of the random variable X. 

The main idea of JADE is to approximate the characteristic function using a given set of moments, and then estimate the PDF $f_X(x)$, using the inverse Fourier transform. An essential element of this estimation is the representation of the  $\langle \mathrm{e}^{\mathrm{i}tX} \rangle_X$ that appears in the definition of the characteristic function. The Jacobi--Anger expansion provides a powerful tool for this purpose.

The Jacobi--Anger expansion is defined as follows~\cite{20_Jacobi-anger_expansion}:
\begin{equation}
\label{eq:A1_Jacobi-Anger_def}
\mathrm{e}^{\mathrm{i} t \cos\theta}
=
\sum_{n=-\infty}^{\infty}{\mathrm{i}^n J_n(t) \mathrm{e}^{\mathrm{i} n \theta}},
\end{equation}
where $t$ is real variable. From the identity $J_{-n}(t)=(-1)^n J_n(t)$, the Jacobi--Anger expansion can be rewritten as follows:
\begin{align}
\begin{split}
    \mathrm{e}^{\mathrm{i} t \cos\theta}
    &=
    J_0(t) + 2 \sum_{n=1}^{\infty}{\mathrm{i}^n J_n(t) \cos {(n \theta)}} \\
    &=
    J_0(t) + 2 \sum_{n=1}^{\infty}{\mathrm{i}^n J_n(t) T_n(x)}
    \label{eq:A2_Jacobi-Anger}
    ,
\end{split}
\end{align}
where $x\coloneqq \cos\theta$. By using Eq.~\eqref{eq:A2_Jacobi-Anger}, we can compute the approximated characteristic function, $\widetilde{\varphi}_X(t)$.
\begin{equation*}
\widetilde{\varphi}_X(t) = J_0(t) + 2 \sum_{n=1}^{N}{\mathrm{i}^n J_n(t)\langle T_n(X) \rangle_X}
\end{equation*}  
At this point, the expectation values of the Chebyshev polynomials, $\langle T_n(X)\rangle_X$  can be easily computed as follows:
\begin{equation}
\label{eq:A4_Chebyshev_expect}
\begin{bmatrix}
\langle T_0(X) \rangle_X \\ \langle T_1(X) \rangle_X \\ \langle T_2(X) \rangle_X \\ \vdots \\ \langle T_N(X) \rangle_X \end{bmatrix}
=
\begin{bmatrix}
c_{0,0} & 0 & 0 & \dots & 0 \\
c_{1,0} & c_{1,1} & 0 & \dots & 0 \\
c_{2,0} & c_{2,1} & c_{2,2} & \dots & 0 \\
\vdots &  & \ddots &  & \vdots \\
c_{N,0} & c_{N,1} & c_{N,2} & \dots & c_{N,N} \\
\end{bmatrix}
\begin{bmatrix}
\mu'_0 \\ \mu'_1 \\ \mu'_2 \\ \vdots \\ \mu'_N \end{bmatrix},
\end{equation}
where $n$, $m$ are non-negative integers and $c_{n,m}$ denotes the coefficient of $x^{m}$ in the Chebyshev polynomial $T_n(x)$. Moreover, since $\left\vert T_n(x) \right\vert \leq 1$ for $x \in [-1, 1] $, both the $\langle T_n(X) \rangle_X$ and $\mu'_n$ must lie within $[-1, 1]$.

\subsection{\texorpdfstring
  {Analytic Fourier transform of\\ the Bessel function of the first kind}
  {Analytic Fourier transform of the Bessel function of the first kind}}
  
\label{sec:Ap_A_2}
The Fourier transform of the Bessel function of the first kind, $J_n(t)$, is expressed as:
\begin{equation}
\label{eq:A5_Bessel_FT}
\frac{1}{\sqrt{2\pi}} \int_{-\infty}^{\infty} J_n(t) \ \mathrm{e}^{\mathrm{i}tx} \ \mathrm{d}t
= \frac{\sqrt{\frac{2}{\pi}} \, \mathrm{i}^n \, \Pi\left(\frac{x}{2}\right) T_n(x)}{\sqrt{1 - x^2}},
\end{equation}
where $n$ is a non-negative integer, and $\Pi(x)$ refers to the rectangular function, which is defined as follows:
\begin{equation*}
\Pi(x)
=
\begin{cases}
0 & \text{if } \; \vert x \vert > \cfrac{1}{2}
\\[6pt] \cfrac{1}{2} & \text{if } \; \vert x \vert = \cfrac{1}{2}
\\[6pt] 1 & \text{if } \; \vert x \vert < \cfrac{1}{2}
\end{cases} \;.
\end{equation*}
Thus, the expression can be rearranged as follows:
\begin{equation}
\label{eq:A6_Bessel_IFT}
\mathcal{F}_t\{J_n(t)\}(-x)
= \int_{-\infty}^{\infty} J_n(t)\,\mathrm{e}^{-\,\mathrm{i} t x}\,\mathrm{d}t
= \frac{2 \mathrm{i}^n \, T_n(-x)}{\sqrt{1 - x^2}},
\end{equation}
where \(\mathcal{F}_t\{g(t)\}(x)\) denotes the Fourier transform of \(g(t)\),
defined by
\begin{equation*}
   \mathcal{F}_t\{g(t)\}(x)
   = \int_{-\infty}^{\infty} g(t)\,\mathrm{e}^{\mathrm{i}xt}\,\mathrm{d}t.  
\end{equation*}

\subsection{Closed form expression for JADE}
\label{sec:Ap_A_3}
As previously described, we estimate the PDF $f_X(x)$ by applying the inverse Fourier transform to the approximated characteristic function, $\widetilde{\varphi}_X(t)$. This can be expressed as follows:
\begin{equation}
\label{eq:A7_JADE_closed_summation}
\begin{aligned}
& f_X(x) \approx \frac{1}{2\pi} \int_{-\infty}^{\infty}{\mathrm{e}^{-\mathrm{i}tx} \ \widetilde{\varphi}_X(t)\, \mathrm{d}t} \\
&\quad = \frac{1}{2\pi} \int_{-\infty}^{\infty}{\mathrm{e}^{- \mathrm{i}tx} \Bigl[J_0(t) + 2 \sum_{n=1}^{N} \mathrm{i}^n J_n(t)\,\langle T_n(X)\rangle_X\Bigr] \, \mathrm{d}t}.
\end{aligned}
\end{equation}
Applying the inverse Fourier transform to each Bessel function of the first kind of order $n$ and summing the results can be reformulated via Eq.~\eqref{eq:A6_Bessel_IFT}.
\begin{equation}
\label{eq:7_JADE_closed}
    f_X(x) \approx
    \frac{1}{\pi}
    \Bigg[
    \frac{\langle T_0(X) \rangle_X T_0(x)}{\sqrt{1-x^2}}
    +
    \sum_{n=1}^N
    \frac{2\langle T_n(X) \rangle_X T_n(x)}{\sqrt{1-x^2}}
    \Bigg]
\end{equation}
Eq.~\eqref{eq:7_JADE_closed} provides a compact and clear representation of JADE. By simply evaluating Eq.~\eqref{eq:7_JADE_closed}, one can successfully estimate the PDF without further iterative procedures or optimization.

\section{Function approximation}
\label{sec:Ap_C}
In this section, we prove that the JADE approximation, as presented in Eq.~\eqref{eq:7_JADE_closed}, is mathematically equivalent to the optimal solution that minimizes the weighted $L_2$ loss. To demonstrate this, we first define the mathematical framework. We work within the weighted space $L^2_w$ on the interval $[-1, 1]$, defined by the weight function $w(x) := \sqrt{1-x^2}$. This space is equipped with the inner product $\langle \cdot, \cdot \rangle_w$ defined as:
\begin{equation}
    \langle F, G \rangle_w := \int_{-1}^{1} F(x)\overline{G(x)}w(x)\mathrm{d}x,
    \label{eq:B_inner_product}
\end{equation}
and the corresponding weighted norm, denoted by $\| \cdot \|_{L^2_w}$, is induced by this inner product as $\|F\|_{L^2_w} := \sqrt{\langle F, F \rangle_w}$.

The JADE approximates a function using a set of basis functions $\{B_n(x)\}_{n=0}^N$, where $B_n(x) := \cfrac{T_n(x)}{\sqrt{1-x^2}}$. A key property of this basis is its orthogonality with respect to this inner product, given by the relation:

\begin{align}
    \nonumber
    \langle B_n, B_m\rangle_{w}
    &=
    \int_{-1}^{1}\!\frac{T_n(x)}{\sqrt{1-x^2}}\,
    \frac{T_m(x)}{\sqrt{1-x^2}}\;\sqrt{1-x^2}\,\mathrm{d}x
    \\
    &=
    \begin{cases}
    0,& n\neq m,\\[2pt]
    \pi,& n=m=0,\\[2pt]
    \frac{\pi}{2},& n=m\neq 0~.
    \end{cases}
    \label{eq:Basis_orthogonality}
\end{align}

Given this framework, the optimal approximation of a target distribution $f_X(x)$ is found by determining the vector of coefficients $\mathbf{t}^* = (t_0^*, t_1^*, \dots, t_N^*)^\top$ that minimizes the squared weighted norm of the residual error. This optimization problem is stated as finding the argument of the minimum:
\begin{equation}
    \mathbf{t}^* = \operatorname*{arg\,min}_{\mathbf t\in\mathbb R^{N+1}} \left\| f_X(x) - \sum_{n=0}^{N} t_n B_n(x) \right\|_{L^2_w}^2.
    \label{eq:B_argmin}
\end{equation}

Because the basis $\{B_n\}$ is orthogonal, the solution to this minimization problem is given by the orthogonal projection of $f_X(x)$ onto the subspace spanned by the basis. The optimal coefficients $t_k^*$ are thus given by:
\begin{align}
    \nonumber
    t^*_k
    &= \cfrac{\langle f_X \,, B_k \rangle_w}
    {\langle B_k \,, B_k \rangle_w}
    =
    \cfrac{\langle T_k(X) \rangle_X}{\displaystyle \int_{-1} ^{1} {\Big [ T_k(x) \Big]^2} \cfrac{\mathrm{d}x}{\sqrt{1-x^2}}}\\ \nonumber
    \\
    &=
    \begin{cases}
    \cfrac{\langle T_k(X) \rangle_X}{\pi} & \text{if } \quad k=0,
    \\[6pt] \cfrac{2\langle T_k(X) \rangle_X}{\pi} & \text{if } \quad k \neq 0,
    \end{cases} \quad
\end{align}

These coefficients are precisely those used in JADE expression (Eq.~\eqref{eq:7_JADE_closed}). From the perspective of polynomial approximation, this proves that this expression coincides with the optimal solution that minimizes the specified weighted $L_2$ loss.

\end{document}